\documentclass[english]{IEEEtran}
\usepackage[latin9]{inputenc}
\usepackage{amsmath}
\usepackage{amssymb}
\usepackage{stackrel}

\makeatletter
\usepackage{algorithm}
\usepackage{algorithmic}

\makeatother

\usepackage{babel}
\begin{document}
\title{A New Derivation for Gaussian Mixture Model Parameter Estimation:
MM Based Approach}
\author{Nitesh Sahu, Prabhu Babu}
\maketitle
\begin{abstract}
In this letter, we revisit the problem of maximum likelihood estimation
(MLE) of parameters of Gaussian Mixture Model (GMM) and show a new
derivation for its parameters. The new derivation, unlike the classical
approach employing the technique of expectation-maximization (EM),
is straightforward and doesn't invoke any hidden or latent variables
and calculation of the conditional density function. The new derivation
is based on the approach of minorization-maximization and involves
finding a tighter lower bound of the log-likelihood criterion. The
update steps of the parameters, obtained via the new derivation, are
same as the update steps obtained via the classical EM algorithm.
\end{abstract}

\begin{IEEEkeywords}
Gaussian mixture model (GMM), Minorization-maximization (MM), Maximum-likelihood
estimation (MLE).
\end{IEEEkeywords}

\section{Introduction}

In the field of machine learning, pattern classification and many
area of statistics, one of the pivotal problem is to estimate the
density or distribution function of the observed data samples. In
the parametric approach of density estimation, a parametric model
for the distribution is assumed, and then the parameters of the model
are determined using the observed finite record of data. A standard
approach to estimate the parameters of the parametric model, from
the given data samples, is maximum likelihood estimation (MLE). In
practice, it is not always possible to describe the structure (distribution)
of observed real-life data samples using a single distribution. To
describe the complex structure in real-life data sets, a linear combination
of several basic distributions is considered as the parametric distribution
model for the data samples, known as mixture model (density). When
the component distributions involved in a mixture model are Gaussian
then the mixture model is called as Gaussian mixture model (GMM).
Superposition of several component distributions can realize the complex
density functions which is not possible with single distribution.
GMMs are also widely used to find underlying clusters in data samples
\cite{bishop2006pattern}.

In this letter, we revisit the parameter estimation problem for GMM
using the minorization-maximization (MM) approach which does not require
the introduction of latent (or hidden) variables and computation of
the conditional expectations which are however essential in classical
expectation-maximization (EM) algorithm based GMM parameter estimation.
The MM based approach of estimating the parameters of GMM is simple
and straightforward to understand. The proposed new derivation, based
on MM approach, produces the same parameter update expressions as
those in the EM algorithm.

\setlength{\belowdisplayskip}{0pt} \setlength{\belowdisplayshortskip}{0pt}

\setlength{\abovedisplayskip}{0pt} \setlength{\abovedisplayshortskip}{0pt}

\section{Problem Formulation\label{sec:Problem-Formulation}}

In this section, we formulate the maximum likelihood parameter estimation
problem for GMM. A general mixture model is described by superposition
of $K$ basic distribution and can be written as

\begin{equation}
p\left(\mathbf{x};\boldsymbol{\theta}\right)=\sum_{k=1}^{K}\pi_{k}f\left(\mathbf{x};\boldsymbol{\theta}_{k}\right)\label{eq:MixMod}
\end{equation}
where each density function $f\left(\mathbf{x};\boldsymbol{\theta}_{k}\right)$,
described by the parameter $\boldsymbol{\theta}_{k}$, is called the
$k^{th}$ mixture component of the mixture density $p\left(\mathbf{x};\boldsymbol{\theta}\right)$,
and $\pi_{k}$ is called the mixing coefficient or proportion. We
denote the parameters associated with mixture density by $\boldsymbol{\theta}$
and $\boldsymbol{\theta}\triangleq\left\{ \left\{ \boldsymbol{\theta}_{k}\right\} _{k=1}^{K},\left\{ \pi_{k}\right\} _{k=1}^{K}\right\} \in\boldsymbol{\Theta}$
where $\boldsymbol{\Theta}$ is parameter space. In order for $p\left(\mathbf{x};\boldsymbol{\theta}\right)$
to qualify as density function $\pi_{k}$ must satisfy $\pi_{k}\geq0\;\forall k$
and $\sum_{k=1}^{K}\pi_{k}=1$.  

The most widely used model to describe the distribution of data samples
is Gaussian distribution which is given by

\begin{equation}
\mathcal{N}\left(\mathbf{x};\boldsymbol{\mu},\mathbf{\Sigma}\right)\triangleq\frac{1}{\sqrt{\left(2\pi\right)^{d}\left|\mathbf{\Sigma}\right|}}\exp\left(-\frac{1}{2}\left(\mathbf{x}-\boldsymbol{\mu}\right)^{T}\mathbf{\Sigma}^{-1}\left(\mathbf{x}-\boldsymbol{\mu}\right)\right)\label{eq:GaussDist}
\end{equation}
where $\mathbf{x}\in\mathbb{R}^{d\times1}$ is data sample, $\boldsymbol{\mu}\in\mathbb{R}^{d\times1}$
denotes the mean and $\mathbf{\Sigma}\succ\boldsymbol{0}$ represents
the covariance matrix. When each component density function $f\left(\mathbf{x};\boldsymbol{\theta}_{k}\right)$
in mixture model (\ref{eq:MixMod}) is multivariate Gaussian then
it is called Gaussian mixture model (GMM). Therefore, GMM can be written
as

\begin{equation}
p\left(\mathbf{x};\boldsymbol{\theta}\right)=\sum_{k=1}^{K}\pi_{k}\mathcal{N}\left(\mathbf{x};\boldsymbol{\mu}_{k},\mathbf{\Sigma}_{k}\right)\label{eq:GMM}
\end{equation}
where $\boldsymbol{\theta}\triangleq\left\{ \left\{ \pi_{k}\right\} _{k=1}^{K},\left\{ \boldsymbol{\mu}_{k},\mathbf{\Sigma}_{k}\right\} _{k=1}^{K}\right\} $.
If the number of component mixture densities is large enough then
GMM can approximate almost any distribution defined on $\mathbb{R}^{d\times1}$
\cite{murphy2012machine}.

Given a data set $\mathbf{\mathcal{D}}=\left\{ \mathbf{x}_{1},\ldots,\mathbf{x}_{N}\right\} $
of $N$ samples generated independently and identically from GMM given
in (\ref{eq:GMM}). The problem is to estimate the parameter $\boldsymbol{\theta}$
using data set $\mathbf{\mathcal{D}}$. Before proceeding, for clarity
of presentation let us define the following function

\begin{equation}
g_{ik}\left(\boldsymbol{\phi}_{k}\right)\triangleq\log\left(\pi_{k}\mathcal{N}\left(\mathbf{x}_{i};\boldsymbol{\mu}_{k},\mathbf{\Sigma}_{k}\right)\right),\label{eq:Sec1Eq6}
\end{equation}
where $\boldsymbol{\phi}_{k}\triangleq\left\{ \pi_{k},\boldsymbol{\mu}_{k},\mathbf{\Sigma}_{k}\right\} $,
which will be used later often. After some manipulation, (\ref{eq:Sec1Eq6})
can be written as

\begin{equation}
\begin{array}{c}
g_{ik}\left(\boldsymbol{\phi}_{k}\right)=\log\pi_{k}-\frac{1}{2}\left(\mathbf{x}_{i}-\boldsymbol{\mu}_{k}\right)^{T}\mathbf{\Sigma}_{k}^{-1}\left(\mathbf{x}_{i}-\boldsymbol{\mu}_{k}\right)\\
-\frac{1}{2}\log\left|\mathbf{\Sigma}_{k}\right|+c,
\end{array}\label{eq:Sec1Eq7}
\end{equation}
where $c\triangleq-\frac{d}{2}\log\left(2\pi\right)$. Thus, from
(\ref{eq:Sec1Eq6}) we have

\begin{equation}
\pi_{k}\mathcal{N}\left(\mathbf{x}_{i};\boldsymbol{\mu}_{k},\mathbf{\Sigma}_{k}\right)=e^{g_{ik}\left(\boldsymbol{\phi}_{k}\right)}.\label{eq:IntroEq7}
\end{equation}

For the data set $\mathbf{\mathcal{D}}$ we can write the likelihood
function as

\begin{equation}
L\left(\boldsymbol{\theta};\mathbf{\mathcal{D}}\right)\triangleq\prod_{i=1}^{N}p\left(\mathbf{x}_{i};\boldsymbol{\theta}\right).\label{eq:LikelihoodFun}
\end{equation}
In MLE, the likelihood function is maximized to estimate the parameters
of the model. Instead of maximizing $L\left(\boldsymbol{\theta};\mathbf{\mathcal{D}}\right)$,
it is more convenient to maximize the logarithm of likelihood function
called the log-likelihood denoted as $l\left(\boldsymbol{\theta};\mathbf{\mathcal{D}}\right)$,
and using (\ref{eq:GMM}), (\ref{eq:IntroEq7}), and (\ref{eq:LikelihoodFun}),
$l\left(\boldsymbol{\theta};\mathbf{\mathcal{D}}\right)$ can be written
as

\begin{equation}
l\left(\boldsymbol{\theta};\mathbf{\mathcal{D}}\right)\triangleq\log L\left(\boldsymbol{\theta};\mathbf{\mathcal{D}}\right)=\sum_{i=1}^{N}\log\left(\sum_{k=1}^{K}e^{g_{ik}\left(\boldsymbol{\phi}_{k}\right)}\right)\label{eq:LogLikelihoodFun}
\end{equation}
where $\boldsymbol{\theta}$ is related to $\boldsymbol{\phi}_{k}$
via $\boldsymbol{\theta}=\left\{ \boldsymbol{\phi}_{k}\right\} _{k=1}^{K}$.

Since logarithm is monotonic function, therefore, the problem of estimating
$\boldsymbol{\theta}$ can be formulated as:

\begin{equation}
\begin{array}{cl}
{\underset{\left\{ \pi_{k},\boldsymbol{\mu}_{k},\mathbf{\Sigma}_{k}\right\} }{\operatorname{maximize}}} & l\left(\boldsymbol{\theta};\mathbf{\mathcal{D}}\right)\\
{\text{ subject to }} & {\boldsymbol{\pi}^{T}\boldsymbol{1}=1,\boldsymbol{\pi}\succeq0,\mathbf{\Sigma}_{k}\succ0\;\forall k}
\end{array}\label{eq:Sec1Eq12}
\end{equation}

The problem in (\ref{eq:Sec1Eq12}) is non-convex as the objective
is a not concave function in the parameters of interest $\boldsymbol{\theta}$.
Moreover, no closed form solution is available for the problem (\ref{eq:Sec1Eq12}).
In the next section, we will see how expectation maximization algorithm
can be applied to arrive at a local maximizer of (\ref{eq:Sec1Eq12}).

\section{Expectation Maximization (EM) Algorithm\label{sec:EM_algo}}

In machine learning and statistics, maximum likelihood (ML) or maximum
a posteriori (MAP) estimate of parameters is easy when complete data
is available. However, when some data is missing and/or model involves
the latent or hidden variables then estimation of parameters becomes
hard \cite{murphy2012machine}. The EM algorithm \cite{dempster1977maximum},
\cite{redner1984mixture} is an iterative method to find the maximum
likelihood estimation of parameters of latent variable models (statistical
models which involve the latent or hidden variable). EM algorithm
alternates between two steps: expectation (E) step and maximization
(M) step. In E-step, conditional expectation of log-likelihood function
is computed given the current estimate of parameters and in M-step,
parameters are obtained by maximizing the conditional expectation
of log-likelihood function created in E-step \cite{alpaydin2014introduction}.

\subsection{EM for GMM}

In this subsection, we discuss the EM algorithm for GMM. We are given
an observed data set $\mathbf{\mathcal{D}}$, and our goal is to find
the parameters $\boldsymbol{\theta}$ of GMM described in (\ref{eq:GMM})
which model the data best. To find $\boldsymbol{\theta}$, our objective
is to maximize the MLE problem given in (\ref{eq:Sec1Eq12}). The
difficulty in maximizing (\ref{eq:Sec1Eq12}) is due to the presence
of summation inside the logarithm of objective function. On the contrary,
EM algorithm handles this issue by introducing the latent variables
and using the notion of complete data log-likelihood. The following
describes how EM algorithm introduces latent variables in the GMM,
which we feel is not that straightforward and can seem very abstract
to a beginner trying to understand GMM.

Assume that the number of component density, $K$, in the GMM is known.
Let us define a $K-$dimensional binary random variable $\mathbf{z}=\left(\begin{array}{ccc}
z_{1} & \ldots & z_{K}\end{array}\right)^{T}$, that is, each $z_{k}\in\left\{ 0,1\right\} $. The random variable
$\mathbf{z}$ is such that only a specific element would be equal
to $1$ $\left(z_{k}=1\right)$ and other elements are zeros. The
random variable $\mathbf{z}$ can take only $K$ possible values $\left\{ \mathbf{e}_{k}\right\} _{k=1}^{K}$
where $\mathbf{e}_{k}$ denotes the $k^{th}$ column of $K\times K$
identity matrix. Therefore, $\mathbf{z}$ follows a multinomial distribution
over $K$ categories (possible values) and this distribution could
be defined in terms of the mixing coefficients $\left\{ \pi_{k}\right\} _{k=1}^{K}$
in (\ref{eq:GMM}) as prior probabilities, that is, probability of
$\mathbf{z}$ taking value $\mathbf{e}_{k}$ is $\pi_{k}$, $p\left(\mathbf{z}=\mathbf{e}_{k}\right)=\pi_{k}$.
Thus, we can write the distribution of $\mathbf{z}$ as

\begin{equation}
p\left(\mathbf{z}\right)=\prod_{k=1}^{K}\pi_{k}^{z_{k}}.
\end{equation}
Since we have already involved $\left\{ \pi_{k}\right\} _{k=1}^{K}$
to define $p\left(\mathbf{z}\right)$, it is safe to say that the
conditional distribution of $\mathbf{x}$ for a given value of $\mathbf{z}=\mathbf{e}_{k}$
is $\mathcal{N}\left(\mathbf{x};\boldsymbol{\mu}_{k},\mathbf{\Sigma}_{k}\right)$,
that is,

\begin{equation}
p\left(\mathbf{x}\mid\mathbf{z}=\mathbf{e}_{k}\right)=\mathcal{N}\left(\mathbf{x};\boldsymbol{\mu}_{k},\mathbf{\Sigma}_{k}\right)
\end{equation}
which can be written as

\begin{equation}
p\left(\mathbf{x}\mid\mathbf{z}\right)=\prod_{k=1}^{K}\mathcal{N}\left(\mathbf{x};\boldsymbol{\mu}_{k},\mathbf{\Sigma}_{k}\right)^{z_{k}}.
\end{equation}
Thus, the joint distribution of $\mathbf{x}$ and $\mathbf{z}$ would
be

\begin{equation}
\begin{aligned}p\left(\mathbf{x},\mathbf{z}\right) & =p\left(\mathbf{z}\right)p\left(\mathbf{x}\mid\mathbf{z}\right)=\prod_{k=1}^{K}\pi_{k}^{z_{k}}\mathcal{N}\left(\mathbf{x};\boldsymbol{\mu}_{k},\mathbf{\Sigma}_{k}\right)^{z_{k}}\end{aligned}
.\label{eq:JointPdf_x_z}
\end{equation}

The conditional probability of $\mathbf{z}$ given $\mathbf{x}$ as
$p\left(\mathbf{z}=\mathbf{e}_{k}\mid\mathbf{x}\right)$ which can
also be written as $p\left(z_{k}=1\mid\mathbf{x}\right)$, can be
given as:

\begin{equation}
\begin{aligned}p\left(z_{k}=1\mid\mathbf{x}\right) & =p\left(\mathbf{z}=\mathbf{e}_{k}\mid\mathbf{x}\right)\\
 & =\frac{p\left(\mathbf{z}=\mathbf{e}_{k}\right)p\left(\mathbf{x}\mid\mathbf{z}=\mathbf{e}_{k}\right)}{\stackrel[j=1]{K}{\sum}p\left(\mathbf{z}=\mathbf{e}_{j}\right)p\left(\mathbf{x}\mid\mathbf{z}=\mathbf{e}_{j}\right)}\\
 & =\frac{\pi_{k}\mathcal{N}\left(\mathbf{x};\boldsymbol{\mu}_{k},\mathbf{\Sigma}_{k}\right)}{\stackrel[j=1]{K}{\sum}\pi_{j}\mathcal{N}\left(\mathbf{x};\boldsymbol{\mu}_{j},\mathbf{\Sigma}_{j}\right)}
\end{aligned}
\end{equation}

Thus, we have successfully introduced latent variable $\mathbf{z}$
and also defined the joint distribution for $\mathbf{z}$ and $\mathbf{x}$
in (\ref{eq:JointPdf_x_z}) for the GMM model. In the steps above
we have associated a latent variable $\mathbf{z}$ with variable $\mathbf{x}$,
similarly, we can associate latent variable $\mathbf{z}_{i}$ with
every data sample $\mathbf{x}_{i}$. Instead of maximizing the log-likelihood
of the incomplete data set $\mathbf{\mathcal{D}}$, one can look at
maximizing the log-likelihood of the complete data set defined as
$\mathbf{\mathcal{D}}_{c}=\left\{ \left(\mathbf{x}_{i},\mathbf{z}_{i}\right)\right\} _{i=1}^{N}$.
The likelihood of the complete data can be written as

\begin{equation}
\begin{aligned}L_{c}\left(\boldsymbol{\theta};\mathcal{D}_{c}\right) & =\prod_{i=1}^{N}p\left(\mathbf{x}_{i},\mathbf{z}_{i}\right)=\prod_{i=1}^{N}\prod_{k=1}^{K}\pi_{k}^{z_{k}^{i}}\mathcal{N}\left(\mathbf{x}_{i};\boldsymbol{\mu}_{k},\mathbf{\Sigma}_{k}\right)^{z_{k}^{i}}\end{aligned}
\end{equation}
where $z_{k}^{i}$ represents $k^{th}$ element of $\mathbf{z}_{i}$.
Taking the logarithm, we get the complete data log-likelihood as

\begin{equation}
l_{c}\left(\boldsymbol{\theta};\mathcal{D}_{c}\right)=\sum_{i=1}^{N}\sum_{k=1}^{K}z_{k}^{i}\log\left(\pi_{k}\mathcal{N}\left(\mathbf{x}_{i};\boldsymbol{\mu}_{k},\mathbf{\Sigma}_{k}\right)\right).\label{eq:sec3Eq23}
\end{equation}

Now, involving the EM algorithm, which comes in two steps: expectation
(E) step and maximization (M) step. In \textbf{E-step}, conditional
expectation of complete data log-likelihood is computed which is defined
as follows:

\begin{equation}
Q\left(\boldsymbol{\theta}\mid\boldsymbol{\theta}_{t}\right)=E\left[l_{c}\left(\boldsymbol{\theta};\mathbf{\mathcal{D}}_{c}\right)\mid\mathbf{\mathcal{D}},\boldsymbol{\theta}_{t}\right]\label{eq:Estep-1}
\end{equation}
where $Q$ is called the auxiliary function \cite{murphy2012machine},
$t$ indexes the iteration, and $\boldsymbol{\theta}_{t}$ is the
parameter values at current iteration $t$. Therefore using (\ref{eq:sec3Eq23})
in (\ref{eq:Estep-1}) we have

\begin{equation}
\begin{aligned}Q\left(\boldsymbol{\theta}\mid\boldsymbol{\theta}_{t}\right) & =\sum_{i=1}^{N}\sum_{k=1}^{K}E\left[z_{k}^{i}\mid\mathbf{\mathcal{D}},\boldsymbol{\theta}_{t}\right]\log\left(\pi_{k}\mathcal{N}\left(\mathbf{x}_{i};\boldsymbol{\mu}_{k},\mathbf{\Sigma}_{k}\right)\right)\\
 & =\sum_{i=1}^{N}\sum_{k=1}^{K}p\left(z_{k}^{i}=1\mid\mathbf{x}_{i},\boldsymbol{\theta}_{t}\right)\log\left(\pi_{k}\mathcal{N}\left(\mathbf{x}_{i};\boldsymbol{\mu}_{k},\mathbf{\Sigma}_{k}\right)\right)\\
 & =\sum_{i=1}^{N}\sum_{k=1}^{K}\gamma_{ik}^{t}\log\left(\pi_{k}\mathcal{N}\left(\mathbf{x}_{i};\boldsymbol{\mu}_{k},\mathbf{\Sigma}_{k}\right)\right)
\end{aligned}
\end{equation}
where

\begin{equation}
\gamma_{ik}^{t}\triangleq\frac{\pi_{k}^{t}\mathcal{N}\left(\mathbf{x}_{i};\boldsymbol{\mu}_{k},\mathbf{\Sigma}_{k}^{t}\right)}{\stackrel[j=1]{K}{\sum}\pi_{j}^{t}\mathcal{N}\left(\mathbf{x}_{i};\boldsymbol{\mu}_{j}^{t},\mathbf{\Sigma}_{j}^{t}\right)}.\label{eq:Sec3Eq28_1}
\end{equation}
Since $z_{k}^{i}$ is binary random variable, that is, $z_{k}^{i}\in\left\{ 0,1\right\} $,
therefore, $E\left[z_{k}^{i}\mid\mathbf{\mathcal{D}},\boldsymbol{\theta}_{t}\right]=p\left(z_{k}^{i}=1\mid\mathbf{x}_{i},\boldsymbol{\theta}_{t}\right)$.

In \textbf{M-step}, parameter updates $\boldsymbol{\theta}_{t+1}$
are obtained by maximizing $Q\left(\boldsymbol{\theta}\mid\boldsymbol{\theta}_{t}\right)$
with respect to $\boldsymbol{\theta}$:

\begin{equation}
\begin{array}{cl}
{\boldsymbol{\theta}_{t+1}=\arg\:\underset{\boldsymbol{\theta}\in\boldsymbol{\Theta}}{\operatorname{maximize}}} & Q\left(\boldsymbol{\theta}\mid\boldsymbol{\theta}_{t}\right)\end{array}.\label{eq:Sec3Eq28}
\end{equation}

In \cite{dempster1977maximum}, it is proved that when $Q\left(\boldsymbol{\theta}\mid\boldsymbol{\theta}_{t}\right)$
increases, the likelihood of the observed data, $l\left(\boldsymbol{\theta};\mathbf{\mathcal{D}}\right)$,
also increases hence a stationary point for $l\left(\boldsymbol{\theta};\mathbf{\mathcal{D}}\right)$
is achieved. Without going into details of solving (\ref{eq:Sec3Eq28}),
which can be referred in detail from \cite{bishop2006pattern}, \cite{murphy2012machine},
the update equations for $\pi_{k}$, $\boldsymbol{\mu}_{k}$ and $\mathbf{\Sigma}_{k}$
are given as \cite{bishop2006pattern}, \cite{murphy2012machine}:

\begin{equation}
\pi_{k}^{t+1}=\frac{\stackrel[i=1]{N}{\sum}\gamma_{ik}^{t}}{N},\label{eq:EM_PI_updtEq}
\end{equation}

\begin{equation}
\boldsymbol{\mu}_{k}^{t+1}=\frac{\stackrel[i=1]{N}{\sum}\gamma_{ik}^{t}\mathbf{x}_{i}}{\stackrel[i=1]{N}{\sum}\gamma_{ik}^{t}},\label{eq:EM_mu_updtEq}
\end{equation}

\begin{equation}
\mathbf{\Sigma}_{k}^{t+1}=\frac{1}{\stackrel[i=1]{N}{\sum}\gamma_{ik}^{t}}\sum_{i=1}^{N}\left(\mathbf{x}_{i}-\boldsymbol{\mu}_{k}^{t+1}\right)\left(\mathbf{x}_{i}-\boldsymbol{\mu}_{k}^{t+1}\right)^{T}.\label{eq:EM_sig_updtEq}
\end{equation}

\section{MM Procedure\label{sec:MM_algo}}

In this section, we briefly describe the MM procedure for a minimization
problem and extension of this idea for a maximization problem is trivial.
Consider the following minimization problem

\begin{equation}
\begin{array}{cl}
{\underset{\mathbf{u}\in\mathcal{U}}{\operatorname{minimize}}} & f\left(\mathbf{u}\right)\end{array}
\end{equation}
where $\mathbf{u}$ is variable and $\mathcal{U}$ is constraint set.

Let $\mathbf{u}_{t}$ denote the estimate of $\mathbf{u}$ at $t-$th
step of MM procedure. A surrogate function $g_{f}\left(\mathbf{u}\mid\mathbf{u}_{t}\right)$
is said to majorize the objective function $f\left(\mathbf{u}\right)$
at $\mathbf{u}_{t}$ if \cite{sun2017majorization}, \cite{hunter2004tutorial}:

\begin{equation}
f\left(\mathbf{u}\right)\leq g_{f}\left(\mathbf{u}\mid\mathbf{u}_{t}\right)\:\forall\mathbf{u}\in\mathcal{U}\label{eq:MM_IneqProp}
\end{equation}
and
\begin{equation}
f\left(\mathbf{u}_{t}\right)=g_{f}\left(\mathbf{u}_{t}\mid\mathbf{u}_{t}\right).\label{eq:MM_EqProp}
\end{equation}

In minimization step, $g_{f}\left(\mathbf{u}\mid\mathbf{u}_{t}\right)$
is minimized instead of $f\left(\mathbf{u}\right)$ and minimizer
of $g_{f}\left(\mathbf{u}\mid\mathbf{u}_{t}\right)$ becomes the estimate
of $\mathbf{u}$ at $\left(t+1\right)-$th iteration of MM hence $\mathbf{u}_{t+1}$
can be written as

\begin{equation}
\begin{array}{cl}
{\mathbf{u}_{t+1}=\arg\:\underset{\mathbf{u}\in\mathcal{U}}{\operatorname{minimize}}} & g_{f}\left(\mathbf{u}\mid\mathbf{u}_{t}\right)\end{array}.\label{eq:MM_updtEq}
\end{equation}
$\mathbf{u}_{t+1}$ evaluated using (\ref{eq:MM_updtEq}) forces the
original objective to decrease as shown below:
\begin{equation}
f\left(\mathbf{u}_{t+1}\right)\overset{\left(\ref{eq:MM_IneqProp}\right)}{\leq}g_{f}\left(\mathbf{u}_{t+1}\mid\mathbf{u}_{t}\right)\overset{\left(\ref{eq:MM_updtEq}\right)}{\leq}g_{f}\left(\mathbf{u}_{t}\mid\mathbf{u}_{t}\right)\overset{\left(\ref{eq:MM_EqProp}\right)}{=}f\left(\mathbf{u}_{t}\right).
\end{equation}
Therefore, starting with an initial point $\mathbf{u}_{0}\in\mathcal{U}$,
MM procedure generates a sequence $\left\{ \mathbf{u}_{t}\right\} $
which monotonically decreases the objective values. Various techniques
and methods to construct the surrogate function are given in \cite{sun2017majorization},
\cite{lange2000optimization}.

\section{Proposed Derivation using MM Approach\label{sec:ProposedDerivation}}

In this section, we approach the problem in (\ref{eq:Sec1Eq12}) as
a maximization problem and show a straightforward way to construct
a minorizing surrogate function, and show how to arrive at the minimizer
of the surrogate function. The parameter update of this MM-based derivation
would be same as in the case of EM algorithm. However, the MM based
derivation would not involve the introduction of any hidden variable
and computation of conditional expectation. We feel that such a straightforward
derivation for the parameter updates would set things clear to a beginner
who is getting introduced to GMM. Before we move into the actual derivation
we will discuss the log-sum-exp function which would be useful in
the proposed derivation. The log-sum-exp function is defined as \cite{boyd2004convex}:

\begin{equation}
h\left(\mathbf{y}\right)\triangleq\log\left(\sum_{k=1}^{n}e^{y_{k}}\right)
\end{equation}
where $\mathbf{y}=\left(\begin{array}{ccc}
y_{1} & \ldots & y_{n}\end{array}\right)^{T}\in\mathbb{R}^{n\times1}$. The log-sum-exp function $h\left(\mathbf{y}\right)$ is convex on
$\mathbb{R}^{n\times1}$. Since $h\left(\mathbf{y}\right)$ is convex
therefore a tight lower bound for $h\left(\mathbf{y}\right)$ at any
$\mathbf{y}_{t}$ can be obtained by writing the first order Taylor
approximation at $\mathbf{y}_{t}$ as given below:

\begin{equation}
h\left(\mathbf{y}\right)\geq s_{h}\left(\mathbf{y}\mid\mathbf{y}_{t}\right)\triangleq h\left(\mathbf{y}_{t}\right)+\nabla h\left(\mathbf{y}_{t}\right)^{T}\left(\mathbf{y}-\mathbf{y}_{t}\right)\label{eq:Sec5Eq15}
\end{equation}
where $\nabla h\left(\mathbf{y}_{t}\right)$ represents the gradient
of $h\left(\mathbf{y}\right)$ computed at $\mathbf{y}_{t}$ and equality
is achieved at $\mathbf{y}=\mathbf{y}_{t}$, that is, $h\left(\mathbf{y}_{t}\right)=s_{h}\left(\mathbf{y_{t}}\mid\mathbf{y}_{t}\right)$.
The gradient of $h\left(\mathbf{y}\right)$ can be computed as

\begin{equation}
\nabla h\left(\mathbf{y}\right)=\left(\sum_{k=1}^{n}e^{y_{k}}\right)^{-1}\left(\begin{array}{c}
e^{y_{1}}\\
\vdots\\
e^{y_{n}}
\end{array}\right).\label{eq:Sec5Eq16}
\end{equation}

The objective function of problem (\ref{eq:Sec1Eq12}) is:

\begin{equation}
l\left(\boldsymbol{\theta};\mathbf{\mathcal{D}}\right)=\sum_{i=1}^{N}\log\left(\sum_{k=1}^{K}e^{g_{ik}\left(\boldsymbol{\phi}_{k}\right)}\right).
\end{equation}
We observe that the function $l\left(\boldsymbol{\theta};\mathbf{\mathcal{D}}\right)$
is sum of log-sum-exp functions in $g_{ik}\left(\boldsymbol{\phi}_{k}\right)$.
We first compute the surrogate function for $l\left(\boldsymbol{\theta};\mathbf{\mathcal{D}}\right)$
at $\boldsymbol{\theta}_{t}$ which lowerbounds $l\left(\boldsymbol{\theta};\mathbf{\mathcal{D}}\right)$
. Using (\ref{eq:Sec5Eq15}), (\ref{eq:Sec5Eq16}) the lower bound
for $\log\left(\stackrel[k=1]{K}{\sum}e^{g_{ik}\left(\boldsymbol{\phi}_{k}\right)}\right)$
at $\left\{ \boldsymbol{\phi}_{k}^{t}\right\} _{k=1}^{K}$ can be
written as follows:

\begin{equation}
\begin{array}{c}
\log\left(\stackrel[k=1]{K}{\sum}e^{g_{ik}\left(\boldsymbol{\phi}_{k}\right)}\right)\geq\log\left(\stackrel[k=1]{K}{\sum}e^{g_{ik}\left(\boldsymbol{\phi}_{k}^{t}\right)}\right)+\\
\left(\mathbf{w}_{i}^{t}\right)^{T}\left(\left(\begin{array}{c}
g_{i1}\left(\boldsymbol{\phi}_{1}\right)\\
\vdots\\
g_{iK}\left(\boldsymbol{\phi}_{K}\right)
\end{array}\right)-\left(\begin{array}{c}
g_{i1}\left(\boldsymbol{\phi}_{1}^{t}\right)\\
\vdots\\
g_{iK}\left(\boldsymbol{\phi}_{K}^{t}\right)
\end{array}\right)\right)
\end{array}\label{eq:Sec5Eq18}
\end{equation}
 where $\mathbf{w}_{i}^{t}=\left(\begin{array}{ccc}
w_{i1}^{t} & \ldots & w_{iK}^{t}\end{array}\right)^{T}$ and
\begin{equation}
w_{ik}^{t}=\frac{e^{g_{ik}\left(\boldsymbol{\phi}_{k}^{t}\right)}}{\stackrel[j=1]{K}{\sum}e^{g_{ij}\left(\boldsymbol{\phi}_{j}^{t}\right)}}.\label{eq:Sec5Eq45}
\end{equation}

Using (\ref{eq:Sec5Eq18}) the lower bound for $l\left(\boldsymbol{\theta};\mathbf{\mathcal{D}}\right)$
at $\boldsymbol{\theta}=\boldsymbol{\theta}_{t}$, noting $\boldsymbol{\theta}=\left\{ \boldsymbol{\phi}_{k}\right\} _{k=1}^{K}$,
can be written as

\begin{equation}
\begin{aligned}l\left(\boldsymbol{\theta};\mathbf{\mathcal{D}}\right) & \geq\begin{array}{c}
l\left(\boldsymbol{\theta}_{t};\mathbf{\mathcal{D}}\right)+\\
\stackrel[i=1]{N}{\sum}\left(\mathbf{w}_{i}^{t}\right)^{T}\left[\left(\begin{array}{c}
g_{i1}\left(\boldsymbol{\phi}_{1}\right)\\
\vdots\\
g_{iK}\left(\boldsymbol{\phi}_{K}\right)
\end{array}\right)-\left(\begin{array}{c}
g_{i1}\left(\boldsymbol{\phi}_{1}^{t}\right)\\
\vdots\\
g_{iK}\left(\boldsymbol{\phi}_{K}^{t}\right)
\end{array}\right)\right]
\end{array}\\
 & =\sum_{i=1}^{N}\sum_{k=1}^{K}w_{ik}^{t}g_{ik}\left(\boldsymbol{\phi}_{k}\right)+\alpha_{t}\\
 & =s_{l}\left(\boldsymbol{\theta}\mid\boldsymbol{\theta}_{t}\right)+\alpha_{t}
\end{aligned}
\end{equation}
where $\alpha_{t}\triangleq l\left(\boldsymbol{\theta}_{t};\mathbf{\mathcal{D}}\right)-\stackrel[i=1]{N}{\sum}\stackrel[k=1]{K}{\sum}w_{ik}^{t}g_{ik}\left(\boldsymbol{\phi}_{k}^{t}\right)$
is a constant and $s_{l}\left(\boldsymbol{\theta}\mid\boldsymbol{\theta}_{t}\right)\triangleq\stackrel[i=1]{N}{\sum}\stackrel[k=1]{K}{\sum}w_{ik}^{t}g_{ik}\left(\boldsymbol{\phi}_{k}\right)$.
The function $s_{l}\left(\boldsymbol{\theta}\mid\boldsymbol{\theta}_{t}\right)+\alpha_{t}$
is global lower bound for $l\left(\boldsymbol{\theta};\mathbf{\mathcal{D}}\right)$
at $\boldsymbol{\theta}=\boldsymbol{\theta}_{t}$, that is, $l\left(\boldsymbol{\theta};\mathbf{\mathcal{D}}\right)\geq s_{l}\left(\boldsymbol{\theta}\mid\boldsymbol{\theta}_{t}\right)+\alpha_{t}$
and equality is achieved at $\boldsymbol{\theta}=\boldsymbol{\theta}_{t}$.

As per MM principle, we need to maximize the surrogate function $s_{l}\left(\boldsymbol{\theta}\mid\boldsymbol{\theta}_{t}\right)+\alpha_{t}$
in lieu of $l\left(\boldsymbol{\theta};\mathbf{\mathcal{D}}\right)$
to obtain the next update for $\boldsymbol{\theta}$, that is, $\boldsymbol{\theta}_{t+1}$.
Hence, leaving the constant term $\alpha_{t}$, $\boldsymbol{\theta}_{t+1}$
can be written as

\begin{equation}
\begin{array}{cl}
{\boldsymbol{\theta}_{t+1}=\arg\underset{\left\{ \pi_{k},\boldsymbol{\mu}_{k},\mathbf{\Sigma}_{k}\right\} }{\operatorname{maximize}}} & s_{l}\left(\boldsymbol{\theta}\mid\boldsymbol{\theta}_{t}\right)\\
{\text{ subject to }} & {\boldsymbol{\pi}^{T}\boldsymbol{1}=1,\boldsymbol{\pi}\succeq0,\mathbf{\Sigma}_{k}\succ0\:\forall k}
\end{array}.\label{eq:Sec5Eq48}
\end{equation}

Using (\ref{eq:Sec1Eq7}), $s_{l}\left(\boldsymbol{\theta}\mid\boldsymbol{\theta}_{t}\right)$
can be written as

\begin{equation}
s_{l}\left(\boldsymbol{\theta}\mid\boldsymbol{\theta}_{t}\right)=\sum_{i=1}^{N}\sum_{k=1}^{K}w_{ik}^{t}\left(\begin{array}{c}
-\frac{1}{2}\left(\mathbf{x}_{i}-\boldsymbol{\mu}_{k}\right)^{T}\mathbf{\Sigma}_{k}^{-1}\left(\mathbf{x}_{i}-\boldsymbol{\mu}_{k}\right)\\
-\frac{1}{2}\log\left|\mathbf{\Sigma}_{k}\right|+\log\pi_{k}+c
\end{array}\right).
\end{equation}
We notice that $s_{l}\left(\boldsymbol{\theta}\mid\boldsymbol{\theta}_{t}\right)$
is separable in $\pi_{k}$ and $\left\{ \boldsymbol{\mu}_{k},\mathbf{\Sigma}_{k}\right\} $,
therefore, the problem (\ref{eq:Sec5Eq48}) can be maximized separately
as two optimization problems in $\pi_{k}$ and $\left\{ \boldsymbol{\mu}_{k},\mathbf{\Sigma}_{k}\right\} $.
The following problem is optimized to obtain the next update $\pi_{k}^{t+1}$:

\begin{equation}
\begin{array}{cl}
{\underset{\left\{ \pi_{k}\right\} }{\operatorname{maximize}}} & \stackrel[i=1]{N}{\sum}\stackrel[k=1]{K}{\sum}w_{ik}^{t}\log\pi_{k}\\
{\text{ subject to }} & {\boldsymbol{\pi}^{T}\boldsymbol{1}=1,\boldsymbol{\pi}\succeq0}
\end{array}\label{eq:Sec5Eq50}
\end{equation}
and $\pi_{k}^{t+1}$ is given by

\begin{equation}
\pi_{k}^{t+1}=\frac{\stackrel[i=1]{N}{\sum}w_{ik}^{t}}{N}
\end{equation}
which is the similar to the update equation as obtained in (\ref{eq:EM_PI_updtEq}).
Next update $\left\{ \boldsymbol{\mu}_{k}^{t+1},\mathbf{\Sigma}_{k}^{t+1}\right\} $
is obtained by solving the following problem:

\begin{equation}
\begin{array}{cl}
{\underset{\left\{ \boldsymbol{\mu}_{k},\mathbf{\Sigma}_{k}\succ0\right\} }{\operatorname{minimize}}} & \stackrel[i=1]{N}{\sum}\stackrel[k=1]{K}{\sum}w_{ik}^{t}\left(\begin{array}{c}
\frac{1}{2}\log\left|\mathbf{\Sigma}_{k}\right|+\\
\frac{1}{2}\left(\mathbf{x}_{i}-\boldsymbol{\mu}_{k}\right)^{T}\mathbf{\Sigma}_{k}^{-1}\left(\mathbf{x}_{i}-\boldsymbol{\mu}_{k}\right)
\end{array}\right)\end{array},\label{eq:Sec5Eq51}
\end{equation}
and given by

\begin{equation}
\boldsymbol{\mu}_{k}^{t+1}=\frac{\stackrel[i=1]{N}{\sum}w_{ik}^{t}\mathbf{x}_{i}}{\stackrel[i=1]{N}{\sum}w_{ik}^{t}}
\end{equation}
and

\begin{equation}
\mathbf{\Sigma}_{k}^{t+1}=\frac{1}{\stackrel[i=1]{N}{\sum}w_{ik}^{t}}\sum_{i=1}^{N}\left(\mathbf{x}_{i}-\boldsymbol{\mu}_{k}^{t+1}\right)\left(\mathbf{x}_{i}-\boldsymbol{\mu}_{k}^{t+1}\right)^{T}.
\end{equation}

Thus, we observe that MM based approach yields the similar update
expressions for $\pi_{k}^{t+1}$, $\boldsymbol{\mu}_{k}^{t+1}$ and
$\mathbf{\Sigma}_{k}^{t+1}$ as obtained in (\ref{eq:EM_PI_updtEq}),
(\ref{eq:EM_mu_updtEq}) and (\ref{eq:EM_sig_updtEq}).

\section{Conclusion\label{sec:Conclusion}}

In this paper, we have revisited the GMM and proposed a new way to
derive its parameters update expressions using MM procedure. The expression
obtained via MM procedure is exactly same as those obtained using
EM algorithm. The MM based derivation is simple and solves the maximum
likelihood estimation problem directly without introducing latent
variable and computing the conditional expectation.

\bibliographystyle{ieeetr}
\bibliography{GMM_submittedToArxiv}

\end{document}